\documentclass[10pt,conference,letterpaper]{IEEEtran}

\usepackage[hidelinks]{hyperref}
\usepackage{packages}

\glsdisablehyper

\loadglsentries{acronyms.sty}

\graphicspath{ {figures} }

\def\figwidth{.95\linewidth}

\begin{document}

\title{Active TLS Stack Fingerprinting:\\Characterizing TLS Server Deployments at Scale
}

\author{\IEEEauthorblockN{Markus Sosnowski\IEEEauthorrefmark{1}, Johannes Zirngibl\IEEEauthorrefmark{1}, Patrick Sattler\IEEEauthorrefmark{1}, Georg Carle\IEEEauthorrefmark{1},\\ Claas Grohnfeldt\IEEEauthorrefmark{2}, Michele Russo\IEEEauthorrefmark{2}, and Daniele Sgandurra\IEEEauthorrefmark{2}} \IEEEauthorblockA{\IEEEauthorrefmark{1}Chair of Network Architectures and Services, Technical University of Munich, Germany\\
\texttt{\{sosnowski, zirngibl, sattler, carle\}@net.in.tum.de}} \IEEEauthorblockA{\IEEEauthorrefmark{2}AI4Sec, Huawei Technologies Munich, Germany\\
\texttt{\{claas.grohnfeldt, michele.russo1, daniele.sgandurra\}@huawei.com}}
}

\maketitle

\begin{abstract}
Active measurements can be used to collect server characteristics on a large scale.\reviewfix{A.1}
This kind of metadata can help discovering hidden relations and commonalities among server deployments offering new possibilities to cluster and classify them.
As an example, identifying a previously-unknown cybercriminal infrastructures can be a valuable source for cyber-threat intelligence. 
We propose herein an active measurement-based methodology for acquiring \gls{tls} metadata from servers and leverage it for their fingerprinting. 
Our fingerprints capture the characteristic behavior of the \acrshort{tls} stack primarily caused by the implementation, configuration, and hardware support of the underlying server. 
Using an empirical optimization strategy that maximizes information gain from every handshake to minimize measurement costs, we generated 10 general-purpose Client Hellos used as scanning probes to create a large database of \acrshort{tls} configurations used for classifying servers. 
We fingerprinted 28 million servers from the Alexa and Majestic toplists and two \gls{cnc} blocklists over a period of 30 weeks with weekly snapshots as foundation for two long-term case studies:
classification of \acrlong{cdn} and \gls{cnc} servers. 
The proposed methodology shows a precision of more than 99\,\% and enables a stable identification of new servers over time.
This study describes a new opportunity for active measurements to provide valuable insights into the Internet that can be used in security-relevant use cases.
\end{abstract}


\glsresetall

\begin{figure}[b]
\scriptsize
\parbox{\dimexpr\linewidth+1\fboxsep}{%
\vspace{-5ex}
\copyright{} IFIP, 2022. This is the author's version of the work. It is posted here by permission of IFIP for your personal use. Not for redistribution. The definitive version was published in TMA CONFERENCE 2022, \url{https://dl.ifip.org/db/conf/tma/tma2022/tma2022-paper35.pdf}
        }%
\begin{tikzpicture}[remember picture,overlay]%
      	\node[draw, text width=\textwidth] at ($(current page.north)-(0,1cm)$) {This is a preprint. If you cite this paper, please use the \emph{TMA} reference: M. Sosnowski, J. Zirngibl, P. Sattler, G. Carle, C. Grohn- feldt, M. Russo, and D. Sgandurra, \enquote{Active TLS Stack Fingerprinting: Characterizing TLS Server Deployments at Scale}, in \emph{Proc. Network Traffic Measurement and Analysis Conference (TMA)}, Jun. 2022.};
\end{tikzpicture}%
\end{figure}

\section{Introduction}

The Internet is characterized by a high level of complexity and heterogeneity.
The arising metadata can enable novel and promising data mining use cases.\reviewfix{A.1,A.2}
One way to handle the complex data is to represent servers in a simple form, that is, by devising server fingerprints that succinctly represent their main characteristics, such as their implementation or configuration options.
The \gls{tls} is currently the \emph{de facto} standard for encrypted communication on the Internet \cite{labovitz2019internet} and has grown to a complex ecosystem due to continuous development and required backward compatibility~\cite{Kotzias2018}.
Due to this, the protocol inherently provides a variety of meta-information related to client and server capabilities that are exchanged during the initial TLS handshake and that can be used to characterize a server.
In previous work, these metadata have been exploited by multiple passive approaches~\cite{Anderson2020,Husak2015,ja3}; whereas, in this work we advocate the usage of active measurements that allow engaging with any responsive server on a large scale, considering encrypted data, and building a comprehensive data set from a single vantage point.\reviewfix{A.2}

Fingerprinting TLS servers can help to understand, model, and secure the Internet.
If TLS fingerprints are able to indicate a level of trust of an infrastructure, they can be valuable cyber-threat intelligence, especially because the \gls{tls} is increasingly utilized by cybercriminals~\cite{quartertls}.
Possible use-cases are: 
\begin{inparaenum}[\itshape (i)]
\item \acrlongpl{ids} fingerprint servers seen in network flows on-demand and compare results with known malicious fingerprints; 
\item security researchers use fingerprints from Internet-wide measurements to find unknown threats; or
\item regularly monitoring own servers where deviations from a fingerprints baseline can indicate an unintended software change or a malware infection.\reviewfix{E.1}
\end{inparaenum}
A need for this kind of information can be derived from the fact that Internet scanning companies like \emph{censys.io} have started to incorporate JARM~\cite{jarm} into their portfolio (according to their host data definition~\cite{censys_data}).
JARM is an open-source TLS server fingerprinting tool that uses similar data to this work and has just recently emerged.\reviewfix{B.1}

However, no long-term systematic study has yet been conducted to show the applicability and the performance of detection use cases with actively collected TLS fingerprints nor has the effectiveness of their collection been analyzed. 

In this work, we will investigate 
\begin{inparaenum}[\itshape (i)]
\item how to construct a similarity relation among \gls{tls} server deployments,
\item how effective scanning configurations can be found while minimizing the measurement costs, and
\item how applications built with these fingerprints perform on a large scale.
\end{inparaenum}

We propose \gls{title} as an effective method for Internet measurements and provide the following contributions:

\begin{compactenum}[\itshape (i)]
    \item reasoned\reviewfix{A.11} selection of \gls{tls} handshake features for fingerprinting the TLS stack and their encoding in an extendable and shareable format;
    \item methodology for finding or tailoring\reviewfix{E.2} effective \glspl{ch} for fingerprinting use cases and 10 general-purpose \glspl{ch} that maximize complementary information extraction from servers;
	\item long running measurement study covering seven months to validate the methodology, show improvements to the related work JARM\reviewfix{B.1,B.15}, and demonstrate the potential of \gls{tls} fingerprinting based on two case studies to detect \gls{cdn} and \gls{cnc} servers; and
	\item published data and open-source scanner to enable reproducible results and to support the community.\footnote{\url{https://active-tls-fingerprinting.github.io}}
\end{compactenum}

\section{Related Work}
\label{sec:related_work}

The large amount of metadata from \gls{tls} handshakes has been used in multiple passive traffic classification and fingerprinting related works~\cite{Anderson2020,Husak2015,ja3}.
In the context of the \gls{tcp}, fingerprinting with active scans has been successfully used by Refs.~\cite{Greenwald2007,Shamsi2016} and \cite{nmap2009} to detect the \gls{os} on a remote server. 
Similar to our \gls{ch} selection, Greenwald~\etal~\cite{Greenwald2007} used the Entropy from the information theory as metric to minimize the number of probes needed for classification.

To the best of our knowledge, the only related work we can directly compare ourselves to is the JARM tool developed by Althouse~\etal~\cite{jarm}. 
It is a popular open-source tool for TLS server fingerprinting. 
Compared to this work, our tool differs in the selection of \glspl{ch} and the extracted features from the \gls{tls} handshake.
They use 10 custom-defined \glspl{ch} for fingerprinting that \enquote{have been specially crafted to pull out unique responses in TLS servers}~\cite{jarm}. 
In contrast to this work, they\reviewfix{B.2} do not complete the \gls{tls} handshake, only use unencrypted data, and do not consider \gls{tls} alert nor extension data besides the \gls{alpn} protocol.
We will show in \cref{sec:compare_jarm} that JARM can be used for \gls{cnc} server detection; however, the effectiveness can be improved with the data suggested by this paper.\reviewfix{A.4,D.1}

Tools like \texttt{testssl.sh}~\cite{testssl} can collect fingerprintable information about \gls{tls} servers, but need a very large number of requests to collect this information.
We observed \texttt{testssl.sh} to make $100-200$ requests to the same server.
This makes it time expensive and ethically questionable to conduct Internet wide scans. 
Additionally, their focus on the configurable part of the \gls{tls} on servers (\eg supported cipher suites) results in neglecting fingerprintable implementation specific features like the extension order.\reviewfix{D.2}

Chung~\etal~\cite{Chung2018} investigated the usage of the \gls{ocsp} stapling from different webservers and observed Nginx servers, which did not return \gls{ocsp} responses to the first client connecting, appending the information only to consecutive requests.
These implementations did not pre-fetch the information nor wait until the \gls{ca} returned the necessary \gls{ocsp} response they could forward to the client. This means, from the clients point of view, that the presence of the \gls{ocsp} stapling response is non-deterministic.\reviewfix{B.3}
This aligns with our observations of the non-deterministic presence of the Status Request extension because this \gls{tls} extension is currently only used to announce an \gls{ocsp} stapling~\cite{rfc6066} response in the handshake.
Gigis~\etal~\cite{sevenyears} investigated Hypergiants, including \glspl{cdn} over a period of seven years.
They showed the increasing role of servers deployed in \glspl{as} not managed by the \gls{cdn} to influence and localize \gls{cdn} traffic to the user.
Their results align with ours because we were also able to find server deployments outside the networks managed by the \gls{cdn} and they found indicators of reverse proxies that influence the measurement results.
Their results rely on servers correctly offering identification material in the certificates, while this work is more subtle and can identify deployments where this information is deliberately hidden (\eg to detect \gls{cnc} servers).\reviewfix{A.3,D.3}

\section{Methodology}
\label{sec:methodology}

The \gls{tls} protocol family \enquote{is the backbone of secure communication over the Internet} (as introduced in more detail by Holz~\etal~\cite{Holz2020}) and currently the \emph{de facto} standard for encrypted communication~\cite{labovitz2019internet}.\reviewfix{B.6}
This work exploits the \gls{tls} to discover similarity relations among servers by fingerprinting their \gls{sb}. 
A \emph{\gls{sb}} is defined as the totality of the capabilities, the interpretation (deviations from the standard or implementation of undefined parts, such as the order of extensions) and the configuration of the \gls{tls} on a server, which can influence the outcome of the \gls{tls} handshake.
Our work is based on the assumption that every \gls{tls} server has a specific \gls{sb} that depends on the implementation, capabilities, and configuration of the \gls{tls} library, hardware, and application utilizing the \gls{tls}.\reviewfix{B.4}
Identifying these behaviors allows to characterize server deployments either directly or in conjunction with additional data (\eg obtained on the \gls{http} level).\reviewfix{B.9}

\gls{tls} handshakes are initiated by clients, and servers only need to react to the initial handshake request (\eg a server chooses one cipher from a list that was previously proposed by the client).
Therefore, the \gls{sb} we want to fingerprint is not directly revealed by the server; only the reaction to different requests (\ie \glspl{ch}) can be observed.
Using multiple \gls{ch} increases the acquirable knowledge and coverage of the \gls{sb}.
For each \gls{ch}, we collect the TLS version, cipher, and TLS extension data from different types of TLS messages to construct the fingerprint.\reviewfix{A.5}
We only initiate handshakes with TLS versions 1.0 to 1.3 but also store a fingerprint if the server answers with an older version.\reviewfix{B.5}

This work proposes a methodology for capturing a part of this \gls{sb} by sending a fixed number of specifically crafted \glspl{ch} to a server, extract features as string encoded information for each \gls{ch}, as detailed in \cref{sec:def_fps}, and combine these features to a fingerprint according to \cref{sec:combining_fps}. 

\subsection{Features Extracted from TLS Handshakes}
\label{sec:def_fps}

Given a \gls{ch}, we extract features from a single handshake in a textually encoded format to be used for fingerprinting.

Features are the selected version, cipher suite, received alerts and extensions data.
Extensions are extracted as an ordered list of key-value pairs each from the \emph{Server Hello}, \emph{Encrypted Extensions}, \emph{Certificate Request}, \emph{Hello Retry Request} and \emph{Certificate} \gls{tls} messages.
The value is only included for several hand-picked extensions (\cref{tab:tls_extension_values}) or else left empty.
An example of a textually represented fingerprint is
\begin{equation*}
\footnotesize
\underbracket{\texttt{771}}_{\mathclap{\text{\small Version}}} \_
\overbracket{\texttt{1301}}^{\mathclap{\text{\small Cipher}}} \_
\underbracket{\texttt{43.AwQ-51.23}}_{\mathclap{\text{\small Server Hello Extensions}}}\_\overbracket{\texttt{0.-10.AAo\ldots}}^{\mathclap{\text{\small Encrypted Extensions}}}\texttt{\_\_\_}
\underbracket{\texttt{18.}}_{\mathclap{\text{\small Certificate Extensions}}}
\overbracket{\texttt{<40}}^{\mathclap{\text{\small Alerts}}}. 
\end{equation*}

\begin{table}
    \centering
    \caption{\Gls{tls} extensions where the data contained in the extensions are used for \Gls{title}.}
    \label{tab:tls_extension_values}
    \begin{tabular}{S[table-format=2.0]lS[table-format=2.0]l}
    \toprule
    {ID} & Name & {ID} & Name\\
    \midrule
    1 & Max Fragment Length & 19 & Client Certificate Type \\
    7 & Client Authentication & 20 & Server Certificate Type\\
    8 & Server Authentication & 24 & Token Binding\\
	9 & Cert Type &27 & Compress Certificate \\
    10 & Supported Groups &28 & Record Size Limit \\
	11 & EC Point Formats & 43 & Supported Versions\\
	13 & Signature Algorithms & 47 & Certificate Authorities \\
	15 & Heartbeat & 50 & Signature Algorithms Cert\\
	16 & \gls{alpn} &51 & Key Share (only selected group) \\
	\bottomrule
    \end{tabular}
\end{table}

In the subsequent paragraphs, we will discuss the reasons for the inclusion of each feature.
The version, selected cipher suite, and used extensions directly depend on the capabilities and the configuration of the \gls{tls} library used by the server and are, therefore, part of the fingerprint.
The content of each individual extension is more diverse, and its fingerprinting value depends on the actual extension.
Information depending on the current \gls{tls} session, specific server, or \gls{tls} certificate are excluded. 
From an analysis of different \gls{tls} extensions~\cite{rfc8446,rfc6066} and from the observations in our scans, we infer the content of the extensions listed in Table A.\ref{tab:tls_extension_values} as a feature.
An exception to this schema is the Key Share extension, where we remove the session specific part and only keep the selected group used for the Diffie\textendash Hellman key exchange.
The fingerprints are defined in a format that can easily be adapted (\eg to include values specified in future extensions).
Error handling is implementation specific; hence, the \gls{tls} alerts sent by the server are also included.
We do not create a fingerprint if the error was caused by the \gls{tcp} layer.
The current approach cannot differentiate whether the error\reviewfix{D.5} is part of the \gls{sb} or if it was a nondeterministic failure of the \gls{tcp} stack.
We consider the order of extensions as an important and implementation-specific information that we include in our fingerprints.
The presence of the Status Request extension was nondeterministic in our measurements (\cref{sec:analysis_sr}); therefore, we removed this extension from the fingerprints trading the information about the \gls{ocsp} stapling support of a server for results consistency.

\subsection{Fingerprinting with Multiple Requests}
\label{sec:combining_fps}

In this work, data from multiple server responses are combined to construct the \gls{tlsfp} of the \gls{sb} because a single response was not precise enough to provide good results in our experiments.

While a single \gls{ch} reveals only a potentially small request-dependent subset of the information about the target server, multiple request\textendash response pairs allow the collection of complementary information and, thus, a more complete picture about the \gls{sb}.
Increasing the number of \glspl{ch} is a trade--off between learned information and measurement costs.
However, the benefit of sending multiple \glspl{ch} decreases with every additional \glspl{ch} one sends because of the limit to which extent a \gls{sb} can influence the \gls{tls} handshake.
Moreover, the number of \glspl{ch} should be limited based on time, resources, and ethical factors.
Hence, the input set $CH$ of \glspl{ch} is an optimizable parameter influencing the effectiveness of fingerprinting. 
Let $f(s,c)$ return the features from a server $s$ given a specific \gls{ch} $c$ in the format described in Section~\ref{sec:def_fps}, then the server fingerprint is defined as
$$ fp(s) = \bigcup\limits_{c \in CH} (c, fp(s, c)). $$
The features obtained with a single \gls{ch} are only comparable in the context of the same \gls{ch}; hence, the \gls{ch} used to generate each part of the fingerprint must be stored along the combined fingerprints.
We never compared information obtained with different \glspl{ch} because we do not know what combination of parameters in the \gls{ch} has caused the particular response.\reviewfix{B.7}

In conclusion, the number of requests and the design of different \glspl{ch} are crucial parameters that can be optimized to maximize the amount of collectible information while minimizing measurement costs and respecting ethical aspects.
This will be experimentally done in \cref{sec:ch_generation_information}.

\subsection{Active Measurements under Ethical Considerations}
\label{sec:active_measurements}
We have used an active measurement pipeline based on established tools and by following basic ethical principles.

The pipeline takes a list of \glspl{ip}, domains, and \glspl{ch} as input.
The domains are resolved according to the IPv4 and IPv6 addresses with MassDNS\footnote{\url{https://github.com/blechschmidt/massdns}} and a local Unbound\footnote{\url{https://www.nlnetlabs.nl/projects/unbound}} server, resulting in a set of \gls{ip} and domain tuples we call targets.
We fingerprint with multiple \glspl{ch}; thus, the final input is a randomly ordered cross-product of the targets and the \glspl{ch}.
This list is fed to the \gls{tls} scanner based on an implementation of Amann~\etal~\cite{amann_https_2017}, a scanner designed for Internet wide usage.
If a domain name is available for an \gls{ip}, we used it as the \gls{sni}.
We designed a custom \gls{tls} library based on the Golang standard library that allows the definition of arbitrary \glspl{ch} as input for each \gls{tls} connection and to extract required \gls{tls} handshake meta-data.
Both scanner and library are open-sourced~\cite{goscanner}.\reviewfix{B.8}

We reduce the impact on third parties by following the best practices, as described by Durumeric~\etal~\cite{durumeric_zmap_2013}.
Our work does not harm individuals or reveal private data as covered by Refs.~\cite{menloreport} and~\cite{PA16} and focuses on publicly reachable services.
We use rate limiting, maintain a blocklist, use dedicated scan servers with abuse contacts, informative \acrshort{rdns} entries, and websites that inform about our research, and provide contact information for further details or scan exclusion. 
Additionally, because we scan the same target with multiple requests, we limit the interference by spreading the requests over a large time frame\reviewfix{A.8,D.6} (\ie two days in the longitudinal study\reviewfix{D.7}).

\section{Systematic Design of Client Hellos}
\label{sec:ch_generation_information}

The internal mechanism of \gls{tls} servers is a black-box for active scanners.
Without knowledge about the implementation of every \gls{tls} server, it is impossible to find the best method for fingerprinting.
However, more effective fingerprints can be developed by optimizing their distinctiveness.
We propose herein an empiric design of \glspl{ch} by analyzing a large pool of randomly generated candidates to find an optimal subset maximizing a given metric.
We choose the total number of distinguishable servers as metric to find general-purpose \glspl{ch} usable for a wide range of use-cases.
If the use-case is known (\eg detecting \gls{cdn} or \gls{cnc} servers), a different strategy could be to minimize the necessary probes needed for a classification. 
We will later revisit this use-case driven design regarding \gls{cnc} servers in \cref{sec:discussion}.\reviewfix{E.3}
We have open-sourced the general-purpose \glspl{ch} and their generation code as part of our scanner~\cite{fingerprinting_web}.\reviewfix{A.5}

The experiment is designed as follows: 
\begin{inparaenum}[\itshape (i)]
    \item randomly generate \num{5000} \glspl{ch} each from the feature space the used \gls{tls} scanner supports and the complete feature space as defined by IANA~\cite{tlsparams};
    \item distribute a measurement with these \glspl{ch} over the Alexa~\cite{alexa} and Majestic~\cite{majestic} toplists with a maximum of \num{13} \glspl{ch} per server to gain a first impression of good-performing \glspl{ch}; and
    \item select the best-performing \glspl{ch} from the previous measurement and conduct a second measurement of the same targets and fingerprint each target with 50 \glspl{ch}.
\end{inparaenum}
We choose the prime-number \num{13} together with a round-robin algorithm to increase the variation of the different sets of \glspl{ch} sent to a single server. 
The decision to scan with 50 \glspl{ch} per server was a pure trade-off between development speed and data quality (the scan took more than \num{4} days).\reviewfix{A.9}

\begin{figure}
    \centering
    \includegraphics[width=\figwidth{}]{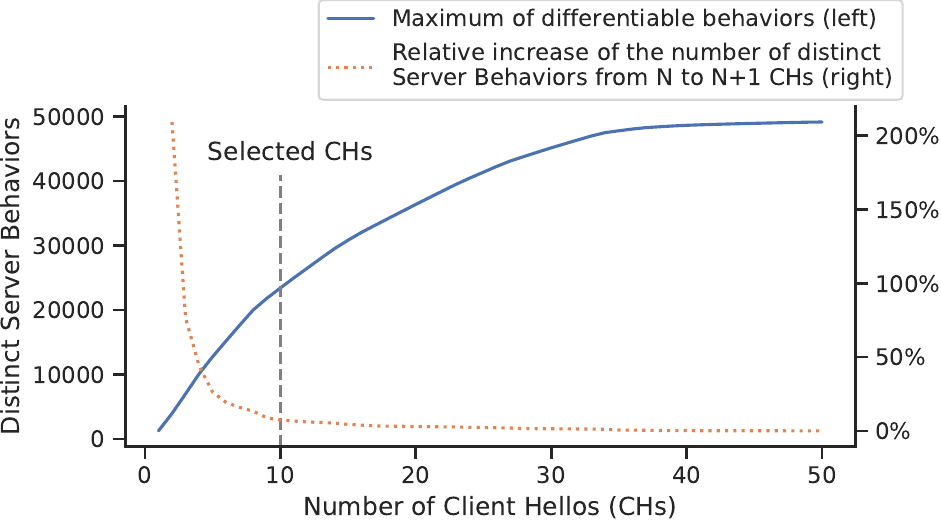}
    \caption{
    Experimentally determining the trade-off to increase the number of \glspl{ch} versus the ability to distinguish \glspl{sb} (measured in distinct fingerprints).
    We selected a set of 10 \glspl{ch} for our following analyses.
    }
    \label{fig:information_50_chs}
\end{figure}

\Cref{fig:information_50_chs} shows the number of behaviors that can be distinguished with subsets of the 50 \glspl{ch}.
The sets were constructed by iteratively selecting \glspl{ch} that increase the number of distinguishable \glspl{sb} most.
To remove the potential bias from nondeterministic \gls{tcp} errors, we consider only servers for which every CH produced a fingerprint.
We can see that every added \gls{ch} enables to differentiate additional \glspl{sb}; however, the relative increase of behaviors decreases the more \glspl{ch} we use. 
We could not reach the theoretical upper limit of distinguishable behaviors.\reviewfix{D.8,B.10}
Based on this analysis, we selected \num{10} general-purpose \glspl{ch} with a good performance in distinguishing \glspl{sb}.
We thought that this number was adequate for our use cases because according to \cref{fig:information_50_chs} the relative increase of information was quite low when using more than 10 \glspl{ch}, we can directly compare related work, and this number seemed to be acceptable for Internet scanners like \texttt{censys.io} already fingerprinting with JARM~\cite{censys_data}. 
However, in \cref{sec:discussion}, we will illustrate that the number can be lower for specific use cases (\eg detecting \gls{cnc} servers).
We only manually adapted some cryptographic parameters of these \glspl{ch} that were too CPU-expensive, such as the 512-bit version (\texttt{secp521r1}~\cite{sec2elliptic}) of the elliptic curve domain parameters for the precomputed \gls{tls} 1.3 Key Share. This curve would have more than doubled our scanning time.

Through the experiment described in this section, we gained a set of 10 general-purpose \glspl{ch} that we will use in the following section to fingerprint servers on the Internet.
They are a good trade-off between limiting the number of requests and the resulting impact on the scanned infrastructure and providing a high distinctiveness of the \glspl{sb}.

\section{Longitudinal Study of Toplists and Blocklists}
\label{sec:analysis_lists}

To investigate the applicability of \gls{tls} fingerprinting on the Internet, we performed measurements of two toplists and two \gls{cnc} blocklists over 7 months.
The following sections analyze the stability of the fingerprints, apply the methodology to detect \gls{cdn} and \gls{cnc} servers, and compare to related work.
The two case studies were selected to have one with a big sample size and where the ground truth can be verified and one where the value of the study is high, but the sample size is low.

\subsection{Data}

We scanned servers from the two toplists and two blocklists over a period of 30 weeks using 25 weekly snapshots starting on July 19, 2021.
Five scans failed and the data for these weeks were skipped.

\Cref{tab:targets_on_lists} presents the number of scanned servers.
A \emph{target} is the scanned combination of \glspl{ip}, \gls{tcp} port, and domain name.
The targets were aggregated over a period of seven days from Alexa~\cite{alexa} and Majestic~\cite{majestic} to cover the weekend effect and the weekly patterns observed by Scheitle~\etal~\cite{scheitle2018toplist}.
The last \num{30} days were used for the SSLBL~\cite{sslbl}, while the current list was utilized for the Feodo Tracker~\cite{feodo}.
We took a larger time frame for the SSLBL, because in addition to the \glspl{ip}, we used the provided certificate hashes to reduce false positives.
This list of around 4M weekly targets was taken as input to the scanning pipeline, as described in \cref{sec:active_measurements}.
The scanning probes are both the 10 \glspl{ch} designed in \cref{sec:ch_generation_information} and the 10 \glspl{ch} modeled after \emph{JARM}~\cite{jarm} to compare both approaches. 
However, unless stated otherwise, the following analyses are only based on the 10 \glspl{ch} designed for this study.
Targets are only considered to be successfully fingerprinted if a fingerprint for each \gls{ch} was collected.
The total number of targets was less than the sum of each list due to an overlap between the lists.

\begin{table}
\centering
\caption{Total number of collected data samples over 30 weeks for the longitudinal analyses. Also shows the number of distinct targets and domain names the data covers ($M = 10^6$).\reviewfix{A.10}}
\label{tab:targets_on_lists}
\begin{tabular}{l r@{~~}r@{~~}r r@{~~}r@{~~}r }
\toprule
\multirow{2}{*}{Source} &     \multicolumn{3}{c}{Scanned} &  \multicolumn{3}{c}{Successful} \\
\cmidrule(lr){2-4} \cmidrule(lr){5-7}
&  {Total}  &  {Targets}  & {Domains} &  {Total} &  {Targets} &  {Domains} \\
\midrule
    Alexa    &  80.88M &   26.48M &   11.69M &        68.48M &          22.21M &           9.87M \\
    Majestic &  37.26M &    4.72M &    1.58M &        31.23M &           3.59M &           1.36M \\
    \sslbl    &   951 &    250 &     &       558   &         127   &            \\
    Feodo    &   8.14k &       1.06k &        &       7.07k   &         883      &               \\
    \midrule
    Total    & 103.77M &   27.69M &   12.01M &        87.87M &          23.12M &          10.16M \\
\bottomrule
\end{tabular}
\end{table}

\subsection{Consistency of TLS Fingerprints}
\label{sec:analysis_sr}

\begin{figure}
    \centering
    \includegraphics[width=\figwidth{}]{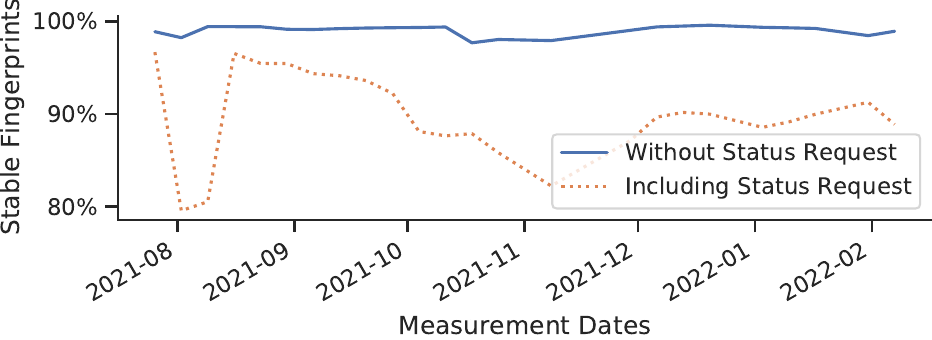}
    \caption{Percentage of targets with the same \gls{tlsfp} on the $n-1$ and $n$th measurement in relation to the total targets fingerprinted on both weeks. The Status Request extensions are responsible for the most unstable fingerprints with drops under 90\% mostly caused by Cloudflare.}
    \label{fig:stability_servers}
\end{figure}

The fingerprints only provide value for identification purposes if they can be unambiguously assigned to a server, and this assignment does not change, in other words, is stable.

For each measurement, a large number of servers was already seen in the last measurement ($\approx 48\%$ each week); hence, their fingerprints can be compared over time.
\Cref{fig:stability_servers} shows the relative number of targets remaining stable during each measurement.
On average the targets remained stable \sperc{99} of the time.
The stability drops on average to \sperc{90} if the Status Request extension is considered for fingerprinting.
In these cases, the presence of the extension is nondeterministic.
This is especially visible during stability drops under \sperc{90}, where only $\approx \frac{3}{4}$ of the \gls{tls} handshakes to Cloudflare contained a Status Request extension compared to the other weeks.

This analysis concludes that Status Request extensions should not be considered for obtaining useful fingerprints.
However, they are, without the extension, a very stable and consistent feature to identify servers. The subsequent sections analyze how differences in fingerprints can be used to identify whole deployments of similar servers.

\subsection{Case study: Detecting CDN server deployments}
\label{sec:analysis_cdn}

A core assumption about \gls{tlsfpi} is that it reveals groups of similar server deployments.
We tested this assumption, by analyzing the fingerprints of four major \glspl{cdn}.
On the one hand, these are \gls{tls}-enabled servers deployed by a single actor on a large scale.
On the other hand, ground truth can be used because it is possible to verify if a server is a part of the \gls{cdn}.
With this analysis, we found servers outside of the \glspl{as} operated by the \gls{cdn} that served the \gls{cdn} content.

The analyzed \glspl{cdn} have in common that they use their own \glspl{as} to deploy servers and \gls{cdn} caches\reviewfix{E.4}. 
Hence, we can identify them through the \gls{as}, determined through the \gls{bgp} dumps downloaded from Routeviews\footnote{\url{https://routeviews.org/}} and Pyasn\footnote{\url{https://pypi.org/project/pyasn/}}.
The content served by a \gls{cdn} is independent of the actual server or the \gls{ip}.
Therefore, we call servers serving this content a \gls{cdn} cache\reviewfix{E.5}.
The \gls{cdn} decides on the criteria like the \gls{sni} which content should be returned. 
In the same manner, the proper \gls{tls} certificate for the requested domain is selected by the \gls{cdn}.
Hence, we can evaluate whether or not a server is a valid \gls{cdn} cache with our TLS scanner.
The server is verified as \gls{cdn} cache if it successfully completes a \gls{tls} handshake for a domain we have manually observed to be cached by the \gls{cdn} and, therefore, proves possession of a valid certificate and the respective private key.

\begin{table}
    \centering
    \caption{Servers seen with a \gls{tlsfp} from the respective \gls{cdn}. 
    A strong correlation between both can be seen.
    }
    \label{tab:cdn_results}
    \begin{threeparttable}
    \begin{tabular}{l S[table-format=5.0] S[table-format=3.0] S[table-format=7.0] S[table-format=6.0]}
        \toprule
        {} &  {Akamai} &  {Alibaba} & {Cloudflare} & {Fastly} \\
        \midrule
        \multicolumn{5}{l}{Targets with a \gls{tlsfp} attributable to a \gls{cdn}} \\
        \midrule
        
        Total            &   98701 &      405 &     7021087 &  572294 \\
        \glspl{ip}        &   24332 &      267 &      222050 &    7238 \\
        \midrule
        \multicolumn{5}{l}{Targets observed from \glspl{as} owned by the \gls{cdn}} \\
        \midrule
        Total              &   97022 &      403 &     6991829 &  559860 \\
        \glspl{ip}  &   23564 &      265 &      214436 &    1752 \\
        \midrule
        \multicolumn{5}{l}{Targets observed from different \glspl{as}, but have access to \gls{cdn} content} \\
        \midrule
        Total     &     664 &         &        2493 &       \\
        \glspl{ip} &     212 &         &         139 &        \\
        \glspl{as}         &      20 &         &          27 &        \\
        \bottomrule
        \end{tabular}
		\end{threeparttable}
\end{table}

This analysis focused on the \glspl{cdn} of Cloudflare, Fastly, Akamai and Alibaba.
The \gls{cdn} fingerprints were collected based on the scanned \glspl{ip} located within their respective \gls{as}.
The \gls{http} Server header is used to enhance the mapping, as described by Gigis~\etal~\cite{sevenyears}.
Note that for Fastly, the \gls{as} was sufficient to detect their \gls{cdn} servers.
\Cref{tab:cdn_results} lists an overview of the results showing a
strong correlation between the \gls{tlsfp} and the \glspl{cdn}.
Almost all targets with a \gls{cdn} fingerprint were located within a \gls{cdn} \gls{as}.
The rate was higher than \sperc{99} for Cloudflare and Fastly.
In addition, \sperc{7} of the targets we falsely assigned to one of the \glspl{cdn} turned out to be valid caches of the \glspl{cdn} from Akamai and Cloudflare outside of their respective networks.
We will discuss these targets more detailed in \cref{sec:discussion}.\reviewfix{B.16}

\begin{figure}
    \centering
    \includegraphics[width=\figwidth{}]{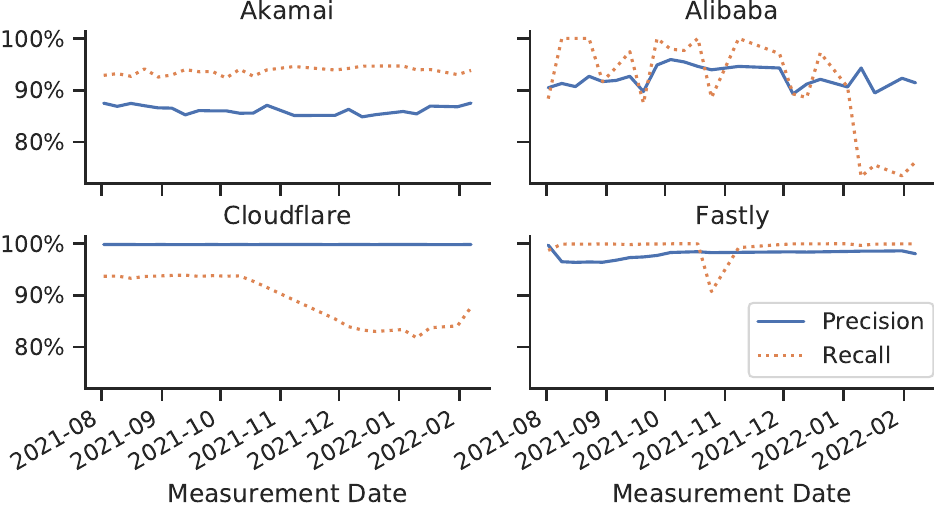}
    \caption{Evaluation of a \gls{cdn} server classifier. It was trained with \glspl{tlsfp} from weeks $[1..n]$ and evaluated on week $n$.}
    \label{fig:cdn_prediction}
\end{figure}

A simple multi-label classifier can be built using these data, by identifying a \gls{cdn} server purely from their \glspl{tlsfp}.
The classifier was trained with the data of weeks $[1..n]$. It classifies a target in the $n+1$th week as a \gls{cdn} server if the fingerprint was observed at least 10 times from the \gls{cdn}, and the certificate was valid in these cases.
This decision was always unambiguous because the fingerprints did not overlap.
We evaluated the metrics precision and recall per \gls{cdn} for each week.
The precision and the recall are defined as $\frac{TP}{TP + FP}$ and $\frac{TP}{PP}$, respectively.
The number of true observations $PP$ is the sum of targets either \first located within the \gls{cdn} \gls{as} with the appropriate \gls{http} Server header values (according to Gigis~\etal~\cite{sevenyears}) or \second validated as a \gls{cdn} cache with the TLS scanner.
\Cref{fig:cdn_prediction} illustrates the results.
The precision was high with more than \sperc{85} for Alibaba and Akamai and more than \sperc{99} for Cloudflare and Fastly.
The networks of the latter were much more uniform and easily clusterable compared to those of Alibaba and Akamai.

An interesting drop of the recall was observed for Fastly in October 2021. This drop was caused by a \gls{sb} change, where new behaviors not covered by fingerprints from the previous measurement dates were observed.
We particularly noticed fewer handshakes to complete with \texttt{http/1.1} as an \gls{alpn}.
The new behavior was seen in the later measurements stabilizing the recall.
Therefore, a potential fingerprint database must be regularly updated to provide the best performance.
We introduced a threshold of 10 valid observations because we had a few cases, in which an Alibaba server responded with a fingerprint we could attribute in most other cases to Tengine webservers (inferred from the HTTP header).
This webserver is not only deployed by Alibaba; thus, it would have produced many false positives.
Not considering these fingerprints caused the recall drop for Alibaba in 2022.
Additionally, Alibaba has a more diverse cloud portfolio on offer compared to the other examined companies that could be the reason for the fluctuations in the precision and recall.\reviewfix{D.10}

Our fingerprints were able to detect minor differences among the deployments of the same \gls{cdn}.
This means, sometimes, the approach was too specific for the general use case to detect just the \gls{cdn}.
We mitigated this problem by mapping multiple fingerprints to each \gls{cdn} covering all of these  variations.
To detect Akamai, Alibaba, Cloudflare, and Fastly we have used \num{86}, \num{1}, \num{801}, and \num{87} fingerprints, respectively.\reviewfix{E.7}

In summary, with \gls{tlsfpi}, large \gls{cdn} deployments can be identified because they share a common \gls{tls} behavior.
The precision was above \sperc{99} for some \glspl{cdn}, and we have found several \gls{cdn} caches in unexpected \glspl{as}.
After showing that the approach works with major known deployments, we will now apply it to a much smaller sample size identifying potentially malicious \gls{cnc} servers.

\subsection{Case study: Identifying C2 servers}

Aside from identifying \gls{cdn} deployments, \gls{tlsfpi} can also be used to identify and track potentially malicious targets like \gls{cnc} servers.

Blocklists containing \gls{cnc} servers are used as an indicator of malicious behavior.
\Cref{tab:cnc_analysis} presents the measurement results for each \gls{cnc} label.
The Ransomware, AsyncRAT, and CobaltStrike labels are from the SSLBL~\cite{sslbl}. The remaining labels are from the Feodo Tracker~\cite{feodo}.
Several fingerprints are unique to a certain type of \gls{cnc} server.
However, these unique fingerprints cover only a small part of all observations.
The number of unique fingerprints gradually increases by combining fingerprints with additional \gls{http} data (\ie the \gls{http} Server header containing values like \texttt{nginx} or \texttt{Apache/2.4.18}).
New servers added to the blocklists repeatedly had the same fingerprint as past servers, that is,
\sperc{95} of the targets added during week $n+1$ have fingerprints already observed for this label during weeks $[1..n]$. 
In other words, these servers could be identified through fingerprinting.

\begin{table}
    \centering
    \caption{Fingerprinting results for the \gls{cnc} servers. Combining our fingerprint (FP) with HTTP data results in more FPs and distinct targets (Tar.) unique to a \gls{cnc} label.}
    \label{tab:cnc_analysis}
    \centering
    \begin{threeparttable}
    \begingroup
    \setlength{\tabcolsep}{2pt}
    \begin{tabular}{l*{3}{S[table-format=3.0]S[table-format=2.0]} S[table-format=3.0] S[table-format=3.0]}
        \toprule
        \multirow{2}{*}{\gls{cnc} Label} & \multicolumn{2}{c}{Total} & \multicolumn{2}{c}{Unique} & \multicolumn{2}{c}{Unique (+HTTP)\tnote{1}} & \multicolumn{2}{c}{New Obs.} \\
        \cmidrule(lr){2-3} \cmidrule(lr){4-5} \cmidrule(lr){6-7} \cmidrule(lr){8-9}
        & {Tar.} & {FPs} & {Tar.} & {FPs} & {Tar.} & {FPs} & {Total} & {Known FP}\\ 
        \midrule
        Dridex&311 &9 & 1 & 2 &137 &4 & 193 &193 \\
        TrickBot&276 & 28 &31 & 8 &106 & 16 & 192 &175 \\
        QakBot&127 &6 & 0 & 0 &122 &3 &86 & 72 \\
        BazarLoader &116 & 16 &27 & 4 & 28 &5 &87 & 40 \\
        Emotet& 92 &3 & 1 & 1 &1 &1 &83 & 62 \\
        
        Ransomware& 12 &2 & 0 & 0 &0 &0 &11 & 10 \\
        CobaltStrike& 11 &9 & 1 & 6 &5 &7 & 4 &4 \\
        AsyncRAT& 11 &3 & 0 & 0 &0 &0 & 9 &7 \\
        \midrule
        Other & 28 & 18 &15 &10 & 16 & 13 &13 &6 \\
        \bottomrule
    \end{tabular}
    \endgroup
    \begin{tablenotes}
		\item [1] Combining \glspl{tlsfp} with the \gls{http} Server header.
	\end{tablenotes}
	\end{threeparttable}
\end{table}

\begin{figure*}
    \centering
    \subfloat[TLS Fingerprint (10 CHs)]{\includegraphics[height=.14\linewidth]{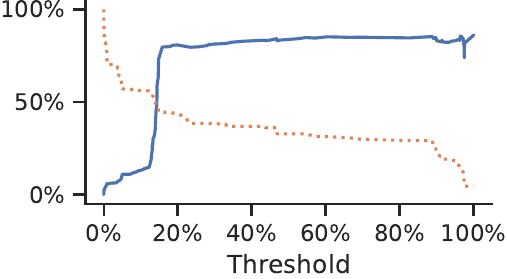}
    \label{fig:cnc_pred_2}}
    \hfil
    \subfloat[HTTP Data (Server Header)]{\includegraphics[height=.14\linewidth]{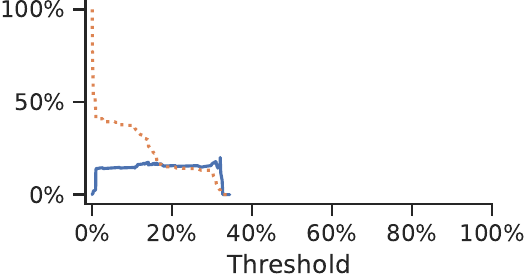}
    \label{fig:cnc_pred_3}}
    \hfil
    \subfloat[Combined (10 CHs + HTTP Data)]{\includegraphics[height=.14\linewidth]{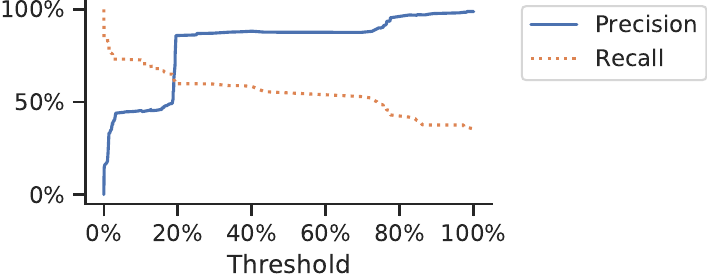}
    \label{fig:cnc_pred_4}}
    \hfil
    \caption{Precision and recall to identify new observations on our input lists as \gls{cnc} servers using the data described in  \cref{fig:cnc_pred_2,fig:cnc_pred_3,fig:cnc_pred_4} as input for the classification.}
    \label{fig:cnc_prediction}
\end{figure*}

This work uses a binary classifier to decide whether or not a server is a \gls{cnc} server from a blocklist.
The classifier takes a threshold $t$ as a parameter and predicts a \gls{cnc} server if the \gls{cnc} rate of the observed fingerprint is higher than $t$. 
This rate is calculated for each fingerprint by dividing the number of times observed from a \gls{cnc} server by the total number of times seen in the training data (\eg a threshold of \sperc{80} means a fingerprint must be observed more than \sperc{80} of the times from blocklists such that servers with this fingerprint are classified as a \gls{cnc} server).
The classifier is evaluated on every new target added to a toplist or a blocklist during week $n+1$ based on the training data from weeks $[1..n]$.
The precision and the recall are defined as $\frac{TP}{TP + FP}$ and $\frac{TP}{PP}$, respectively.
The true observations $PP$ are the responsive targets found on a blocklist.
\Cref{fig:cnc_prediction} shows results for different thresholds for three data sources to construct the input for the classification.
The classifier performance will greatly increase if the fingerprints are combined with additional data from the \gls{http} Server headers.
This \gls{http} datum alone is unsuited for a classifier, but when combined, it achieves a maximum precision of \sperc{99} for \sperc{35} of the added \gls{cnc} servers.
In contrast to the \gls{cdn} detection, the low recall is an indicator that our fingerprinting was not fine-grained enough to detect the differences in the deployments needed to identify all \gls{cnc} servers.
Augmenting the fingerprints with \gls{http} data was our solution to improve this granularity for more effective fingerprinting.\reviewfix{E.3}

This analysis demonstrates that many \gls{cnc} servers have a very unique \gls{tls} behavior that can be used to identify them.
We also presented how a classification of these servers can be implemented on a large scale, and that such an approach can achieve a high precision.
Additionally, a potential fingerprint database for \gls{cnc} servers would live much longer than the entries on the used blocklists, which means that they can provide valuable information about newly deployed \gls{cnc} servers until their \glspl{ip} get publicly known.

\subsection{Comparison With JARM}
\label{sec:compare_jarm}

After analyzing the applicability of \glspl{tlsfpi} on a large scale, the subsequent paragraphs have a look at the performance of JARM~\cite{jarm} which can also be used to fingerprint TLS servers.
While JARM uses similar data to our approach, we will show that both the data extracted from the TLS handshakes and the \gls{ch} selection of this work provide an improved base for fingerprinting.

We scanned every target with the \glspl{ch} used by JARM, as well as the empirically optimized ones from this work. Thus, we can compare both approaches.
We did not use the open-source JARM script directly because it was not able to scan our number of targets on our hardware fast enough.\reviewfix{D.11}
Instead, we have used our own scanner with the JARM \glspl{ch} and extracted\reviewfix{D.12} the subset of features that JARM uses to construct its fingerprints from our data.
In particular, fingerprints were stripped from alerts, any TLS message besides the Server Hello, and any extension data besides the \gls{alpn} (\ie the IDs and the order of the extensions remained intact, and only the data contained in these extensions were removed).
\Cref{tab:comp_jarm} presents a comparison of how the selection of features and \glspl{ch} affects the fingerprinting results.
The improvements proposed herein consistently provide better results while maintaining the number of requests necessary for fingerprinting the same (\ie 10 \glspl{ch}).
In total, this work can differentiate \sperc{55} more \glspl{sb}.
Considering the \gls{cnc} servers, this improved differentiation results in 16 additional unique \gls{cnc} behaviors and four times more \gls{cnc} servers identifiable with these unique behaviors.
In this case, \enquote{unique} means no overlap was observed with any server found on a toplist.

\begin{table}
    \centering
    \caption{Comparing related work JARM with Active TLS Stack fingerprinting (ATSF) considering both the dimensions of feature selection and used \glspl{ch}.}
    \label{tab:comp_jarm}
    \begin{threeparttable}
    \begin{tabular}{l S[table-format=5.0] @{~~} S[table-format=5.0] S[table-format=5.0] @{~~} S[table-format=5.0]}
        \toprule
        
        {Feature selection approach} & \multicolumn{2}{c}{JARM} & \multicolumn{2}{c}{ATSF} \\
        
        \cmidrule(lr){2-3} \cmidrule(lr){4-5}
        Used \glspl{ch} & {JARM} & {ATSF} & {JARM} & {ATSF} \\

        \midrule
        Unique fingerprints           & 52316 & 63037 & 65111 & 81180 \\
        Unique \gls{cnc} fingerprints &  7 & 12 & 16 & 23  \\
        Unique \gls{cnc} targets      &  15 & 43 & 22 & 64 \\
        \bottomrule
    \end{tabular}
	\end{threeparttable}
\end{table}

In conclusion, TLS fingerprinting tools like JARM can benefit from the advanced feature extraction and the systematic design of the \glspl{ch} proposed in this work to improve the effectiveness of the approach.
\section{Discussion}
\label{sec:discussion}

With our fingerprinting approach we gained new insights into the Internet and found interesting relations among \gls{tls} servers.
Some of them, we discuss in the following paragraphs.

\paragraph{Advanced Similarity Comparison}
This work explicitly does not obfuscate any information as done by Refs.~\cite{ja3} and \cite{jarm} to keep the syntactic information of each part of the fingerprint intact.
This supports explainability, allows to relate not only equal but similar behaviors in the future, and to adapt the fingerprints afterwards (\eg removing the Status Request extensions).
Similar fingerprints can indicate deployments from an actor who has done just minor configuration changes.

\paragraph{The Success of Random \glspl{ch}}
In the beginning we have used the standard \gls{ch} from the Go library and \glspl{ch} mimicking popular browsers for fingerprinting.
However, they could not extract enough information from servers to be effective in use cases because the responses were similar focusing on few popular TLS configurations.
In contrast, the Random \glspl{ch} were empirically optimized to distinguish servers and have unusual combinations and order of parameters.
They vary in the combination of TLS versions, ciphers, \glspl{alpn}, and supported groups and,
sometimes, are not realistic (\eg offer \glspl{alpn} unsuitable for webservers).
Interestingly, two \glspl{ch} use TLS 1.3 and none use TLS 1.2 as \texttt{legacy\_version}, both is not conform to the RFC~\cite{rfc8446}.
In contrast, JARM uses more realistic \glspl{ch} with fewer variations in the parameters.
They defined them through systematic subsets (\eg the top half) or a reversed order from a fixed input.
In conclusion, the Random \glspl{ch} are very successful for fingerprinting because they have a fuzzy character triggering more distinctive responses.
\reviewfix{B.11,B.12,B.13,B.14,A.6}

\begin{figure}[t]
    \centering
    \includegraphics[width=\figwidth{}]{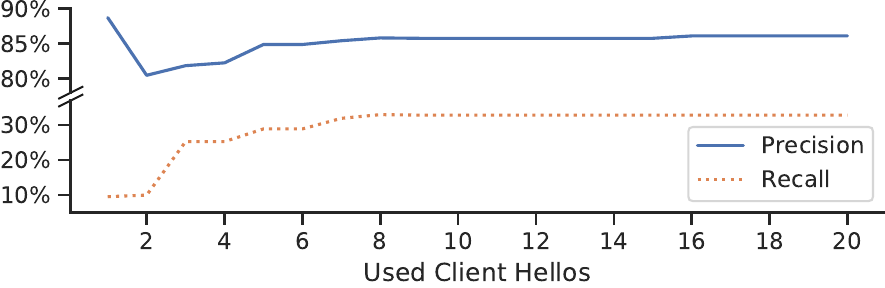}
    \caption{Influence of the \glspl{ch} on the \gls{cnc} detection with an \sperc{80} threshold.}
    \label{fig:ch_c2_prediction}
\end{figure}

\paragraph{Adequate number of \glspl{ch} for \gls{cnc} detection}
\label{par:ch_use_case_driven}
We designed the 10 \glspl{ch} as a general-purpose configuration to provide a good base for a classification.
However, for specific use cases they can be different.
We scanned every server with 10 optimized and 10 JARM \glspl{ch}; hence, the classification performance shown in \cref{fig:cnc_pred_2} can be recomputed using up to 20 \glspl{ch} as input for an \sperc{80} threshold.
We used the same strategy to add \glspl{ch} to the classification input as in \cref{sec:ch_generation_information}.
While the precision was high for almost all sets of \glspl{ch}, scanning with multiple ones mainly increased the number of classified servers (visible in the recall).
The maximum precision and recall was achieved with 16 \glspl{ch}, but the gain was minimal after eight \glspl{ch}. 
For this detection use case, multiple \glspl{ch} are necessary, but eight \glspl{ch} would have been sufficient.
Additionally, an adaptive scanning approach could be implemented in the future work, where additional requests are only sent to a server if the precision of its current classification is not high enough.

\paragraph{\gls{cdn} Caches in a Foreign \gls{as}}
\label{par:dis_offnets}
We were not able to correctly identify all reasons why some domains resolved to \gls{cdn} caches in a foreign \gls{as}.
In particular, \sperc{80} of them were located in an \gls{as} from VTC Digicom, Amazon, or Render.
Interestingly, the whole Render \gls{as} was proxied through Cloudflare and its \gls{ip} prefix functioned as \gls{cdn} cache.
It is possible that we have observed the effect of operators that try to remain in control of their traffic flow (\eg by deploying a Meta-CDN~\cite{metacdn}) because \sperc{68} of the domains pointing to a \gls{cdn} cache in a foreign \gls{as} used a nameserver unrelated to the \gls{cdn}.
To our knowledge, Cloudflare does not operate \gls{cdn} caches outside of their \glspl{as}.
At least six of these caches turned out to be reverse proxies set up by third parties; investigated with a tracing endpoint\footnote{\url{https://cloudflare.com/cdn-cgi/trace}} suggested to us by Cloudflare.
Additionally, our data contained \num{1094} outlier domain names resolving to the public \gls{dns} resolver \texttt{1.1.1.1}, which was apparently also a cache of the Cloudflare network.
\sperc{76} of these domain names used a Cloudflare nameserver.
In contrast to Cloudflare, Akamai has deployed \gls{cdn} caches in more than 1k foreign \glspl{as} to localize their traffic~\cite{Nygren2010}.
However, we detected just $20$ \glspl{as} because we did not scan the full IPv4 address space, but \glspl{ip} resolved from the toplists.
Akamai uses the \gls{dns} to distribute the load on their servers~\cite{Nygren2010}.
We assume, we saw these \glspl{as} because our scan traffic was not always directed to the closest \gls{cdn} cache but distributed across servers in multiple \glspl{as}.\reviewfix{E.6,E.5,C.1}

\paragraph{\gls{cdn} Inconsistencies}
Some \gls{cdn} caches were inconsistent in their responses because not every \gls{ip} successfully responded to every domain name requested. 
To the end, multiple domain names were necessary to validate caches.
For Cloudflare, this was only a single \gls{ip} located in China. 
For Akamai, bigger clusters were visible and multiple domain names were needed to verify them.

\paragraph{Unstable Fingerprints}
Some targets had inconsistent fingerprints that could be caused by the server or by more complex setups.
Sometimes, we saw indicators of load balancers in the \gls{http} Server header indicating that the actual fingerprinted server could change during the scan process.
This is also the main limitation of our approach because it relies on multiple \gls{tls} connections to connect to the same \gls{sb}.
However, this was rarely an issue (\cref{sec:analysis_sr}).

\section{Conclusion}
\label{sec:conclusion}

This work proposed a methodology for acquiring and leveraging \gls{tls} metadata with the help of large scale active measurements.
The value of the approach is backed by two measurement studies on the Alexa and Majestic toplists and two \gls{cnc} blocklists over half a year to detect \gls{cdn} and \gls{cnc} servers. 
New \gls{cnc} servers added to the blocklists were classified with a precision higher than \sperc{99}. 
Depending on the \gls{cdn} and their infrastructure, the detection precision ranged from \sperc{85} (Akamai and Alibaba) to more than \sperc{99} (Cloudflare and Fastly).
Additionally, \num{351} IP addresses were identified serving \gls{cdn} content outside of the \acrshort{as} operated by the \gls{cdn}.

The results were obtained with a reasoned selection of the features extracted from \gls{tls} handshakes and with the use of multiple scanning probes to construct fingerprints of the TLS stack on servers.
These 10 probes were empirically optimized to provide as much information as possible while minimizing the measurement time and the ethical impact on targets.

This paper describes in detail how TLS stack fingerprinting can be efficiently conducted and proves that these data can be applied on real-world classification problems like \gls{cnc} detection to provide valuable security-related insights.

Moreover, the extended feature extraction and improved \gls{ch} design can improve existing \gls{tlsfpi} tools while maintaining the active scanning effort.
Given the approach is independent of the actual \glspl{ch}, we expect future works to tune their \glspl{ch} to specific use cases, or individually adapt them on a per-server level.

\def\UrlBreaks{\do/\do-}

\bibliographystyle{IEEEtran}
\bibliography{IEEEabrv,lit}

\begin{thebibliography}{10}
\providecommand{\url}[1]{#1}
\csname url@samestyle\endcsname
\providecommand{\newblock}{\relax}
\providecommand{\bibinfo}[2]{#2}
\providecommand{\BIBentrySTDinterwordspacing}{\spaceskip=0pt\relax}
\providecommand{\BIBentryALTinterwordstretchfactor}{4}
\providecommand{\BIBentryALTinterwordspacing}{\spaceskip=\fontdimen2\font plus
\BIBentryALTinterwordstretchfactor\fontdimen3\font minus
  \fontdimen4\font\relax}
\providecommand{\BIBforeignlanguage}[2]{{%
\expandafter\ifx\csname l@#1\endcsname\relax
\typeout{** WARNING: IEEEtran.bst: No hyphenation pattern has been}%
\typeout{** loaded for the language `#1'. Using the pattern for}%
\typeout{** the default language instead.}%
\else
\language=\csname l@#1\endcsname
\fi
#2}}
\providecommand{\BIBdecl}{\relax}
\BIBdecl

\bibitem{labovitz2019internet}
C.~Labovitz, ``Internet traffic 2009-2019,'' in \emph{Proc. Asia Pacific
  Regional Internet Conf. Operational Technologies}, 2019.

\bibitem{Kotzias2018}
P.~Kotzias, A.~Razaghpanah, J.~Amann, K.~G. Paterson, N.~Vallina-Rodriguez, and
  J.~Caballero, ``{Coming of Age: A Longitudinal Study of TLS Deployment},'' in
  \emph{Proc. ACM Int. Measurement Conference (IMC)}, 2018.

\bibitem{Anderson2020}
B.~Anderson and D.~A. McGrew, ``Accurate {TLS} fingerprinting using destination
  context and knowledge bases,'' \emph{CoRR}, vol. abs/2009.01939, 2020.

\bibitem{Husak2015}
M.~Husák, M.~Cermák, T.~Jirsík, and P.~Celeda, ``{Network-Based HTTPS Client
  Identification Using SSL/TLS Fingerprinting},'' in \emph{2015 10th
  International Conference on Availability, Reliability and Security}, 2015.

\bibitem{ja3}
\BIBentryALTinterwordspacing
J.~Althouse, J.~Atkinson, and J.~Atkins, ``{TLS Fingerprinting with JA3 and
  JA3S},'' 2019. [Online]. Available:
  \url{https://engineering.salesforce.com/tls-fingerprinting-with-ja3-and-ja3s-247362855967}
\BIBentrySTDinterwordspacing

\bibitem{quartertls}
\BIBentryALTinterwordspacing
{Sean Gallagher}, ``{Nearly half of malware now use TLS to conceal
  communications},'' 2021. [Online]. Available:
  \url{https://news.sophos.com/en-us/2021/04/21/nearly-half-of-malware-now-use-tls-to-conceal-communications/}
\BIBentrySTDinterwordspacing

\bibitem{jarm}
\BIBentryALTinterwordspacing
J.~Althouse, A.~Smart, R.~Nunnally, Jr., and M.~Brady, ``{Easily Identify
  Malicious Servers on the Internet with JARM},'' 2020. [Online]. Available:
  \url{https://engineering.salesforce.com/easily-identify-malicious-servers-on-the-internet-with-jarm-e095edac525a}
\BIBentrySTDinterwordspacing

\bibitem{censys_data}
\BIBentryALTinterwordspacing
{Censys.io}. {Data Definitions}. Last accessed Feb. 24, 2022. [Online].
  Available: \url{https://search.censys.io/search/definitions}
\BIBentrySTDinterwordspacing

\bibitem{Greenwald2007}
L.~G. Greenwald and T.~J. Thomas, ``Toward undetected operating system
  fingerprinting,'' in \emph{Proceedings of the First USENIX Workshop on
  Offensive Technologies}.\hskip 1em plus 0.5em minus 0.4em\relax USENIX
  Association, 2007.

\bibitem{Shamsi2016}
Z.~Shamsi, A.~Nandwani, D.~Leonard, and D.~Loguinov, ``{Hershel: Single-Packet
  OS Fingerprinting},'' \emph{IEEE/ACM Transactions on Networking}, 2016.

\bibitem{nmap2009}
G.~F. Lyon, \emph{Nmap Network Scanning: The Official Nmap Project Guide to
  Network Discovery and Security Scanning}.\hskip 1em plus 0.5em minus
  0.4em\relax Insecure, 2009.

\bibitem{testssl}
\BIBentryALTinterwordspacing
{Dirk Wetter}. {Testing TLS/SSL encryption}. Last accessed March 1, 2022.
  [Online]. Available: \url{https://testssl.sh/}
\BIBentrySTDinterwordspacing

\bibitem{Chung2018}
T.~Chung, J.~Lok, B.~Chandrasekaran, D.~Choffnes, D.~Levin, B.~M. Maggs,
  A.~Mislove, J.~Rula, N.~Sullivan, and C.~Wilson, ``{Is the Web Ready for OCSP
  Must-Staple?}'' in \emph{Proc. ACM Int. Measurement Conference (IMC)}, 2018.

\bibitem{rfc6066}
D.~E. Eastlake, ``{Transport Layer Security (TLS) Extensions: Extension
  Definitions},'' RFC 6066, 2011.

\bibitem{sevenyears}
P.~Gigis, M.~Calder, L.~Manassakis, G.~Nomikos, V.~Kotronis, X.~Dimitropoulos,
  E.~Katz-Bassett, and G.~Smaragdakis, ``{Seven Years in the Life of
  Hypergiants' off-Nets},'' in \emph{Proc. ACM SIGCOMM}, 2021.

\bibitem{Holz2020}
R.~Holz, J.~Hiller, J.~Amann, A.~Razaghpanah, T.~Jost, N.~Vallina-Rodriguez,
  and O.~Hohlfeld, ``{Tracking the Deployment of TLS 1.3 on the Web: A Story of
  Experimentation and Centralization},'' \emph{ACM SIGCOMM Computer
  Communication Review}, 2020.

\bibitem{rfc8446}
E.~Rescorla, ``{The Transport Layer Security (TLS) Protocol Version 1.3},'' RFC
  8446, Aug. 2018.

\bibitem{amann_https_2017}
J.~Amann, O.~Gasser, Q.~Scheitle, L.~Brent, G.~Carle, and R.~Holz, ``{Mission
  Accomplished? HTTPS Security after Diginotar},'' in \emph{Proc. ACM Int.
  Measurement Conference (IMC)}, 2017.

\bibitem{goscanner}
O.~Gasser, M.~Sosnowski, P.~Sattler, and J.~Zirngibl. {Goscanner}.
  \url{https://github.com/tumi8/goscanner}.

\bibitem{durumeric_zmap_2013}
Z.~Durumeric, E.~Wustrow, and J.~A. Halderman, ``{ZMap}: Fast internet-wide
  scanning and its security applications,'' in \emph{Proc. USENIX Security
  Symposium}, 2013.

\bibitem{menloreport}
D.~Dittrich and E.~Kenneally, ``{The Menlo Report: Ethical principles guiding
  information and communication technology research},'' \emph{US Department of
  Homeland Security}, 2012.

\bibitem{PA16}
C.~Partridge and M.~Allman, ``{{Addressing Ethical Considerations in Network
  Measurement Papers}},'' \emph{Communications of the ACM}, 2016.

\bibitem{fingerprinting_web}
\BIBentryALTinterwordspacing
``{Active TLS Stack Fingerprinting: Additional Material"},'' 2022. [Online].
  Available: \url{https://active-tls-fingerprinting.github.io/}
\BIBentrySTDinterwordspacing

\bibitem{tlsparams}
\BIBentryALTinterwordspacing
IANA, ``{Transport Layer Security (TLS) Parameters},'' last accessed March 1,
  2021. [Online]. Available:
  \url{https://www.iana.org/assignments/tls-parameters/tls-parameters.xhtml}
\BIBentrySTDinterwordspacing

\bibitem{alexa}
\BIBentryALTinterwordspacing
{Alexa}, ``{Top 1M sites},'' last accessed 28 Feb. 2022. [Online]. Available:
  \url{http://s3.dualstack.us-east-1.amazonaws.com/alexa-static/top-1m.csv.zip}
\BIBentrySTDinterwordspacing

\bibitem{majestic}
\BIBentryALTinterwordspacing
{Majestic}, ``{The Majestic Million},'' last accessed 28 Feb. 2022. [Online].
  Available: \url{https://majestic.com/reports/majestic-million/}
\BIBentrySTDinterwordspacing

\bibitem{sec2elliptic}
\BIBentryALTinterwordspacing
D.~R.~L. Brown, ``{SEC 2: Recommended Elliptic Curve Domain Parameters},''
  2010. [Online]. Available: \url{http://www.secg.org/sec2-v2.pdf}
\BIBentrySTDinterwordspacing

\bibitem{scheitle2018toplist}
Q.~Scheitle, O.~Hohlfeld, J.~Gamba, J.~Jelten, T.~Zimmermann, S.~D. Strowes,
  and N.~Vallina-Rodriguez, ``{A Long Way to the Top: Significance, Structure,
  and Stability of Internet Top Lists},'' in \emph{Proc. ACM Int. Measurement
  Conference (IMC)}, 2018.

\bibitem{sslbl}
\BIBentryALTinterwordspacing
{abuse.ch}, ``{SSL Certificate Blacklist},'' last accessed Feb. 28, 2022.
  [Online]. Available: \url{https://sslbl.abuse.ch/}
\BIBentrySTDinterwordspacing

\bibitem{feodo}
\BIBentryALTinterwordspacing
------, ``{Feodo Tracker},'' last accessed Feb. 28, 2022. [Online]. Available:
  \url{https://feodotracker.abuse.ch/}
\BIBentrySTDinterwordspacing

\bibitem{metacdn}
O.~Hohlfeld, J.~Rüth, K.~Wolsing, and T.~Zimmermann, ``{Characterizing a
  Meta-CDN},'' in \emph{Proc. Passive and Active Measurement (PAM)}, 2018.

\bibitem{Nygren2010}
E.~Nygren, R.~K. Sitaraman, and J.~Sun, ``{The Akamai Network: A Platform for
  High-Performance Internet Applications},'' \emph{ACM SIGOPS Oper. Syst.
  Rev.}, 2010.

\end{thebibliography}



\end{document}